\documentclass[
showpacs,preprintnumbers,amsmath,amssymb]{revtex4}
\usepackage{amsmath}

\begin{document}

\tolerance=5000

\def\pp{{\, \mid \hskip -1.5mm =}}
\def\cL{{\cal L}}
\def\be{\begin{equation}}
\def\ee{\end{equation}}
\def\bea{\begin{eqnarray}}
\def\eea{\end{eqnarray}}
\def\beq{\begin{eqnarray}}
\def\eeq{\end{eqnarray}}
\def\tr{{\rm tr}\, }
\def\nn{\nonumber \\}
\def\e{{\rm e}}

\title{Newton law corrections and instabilities in $f(R)$ gravity with the
effective cosmological constant epoch}
\author{Shin'ichi Nojiri}
\email{nojiri@phys.nagoya-u.ac.jp}
\affiliation{Department of Physics, Nagoya University, Nagoya 464-8602. Japan}
\author{Sergei D. Odintsov\footnote{also at Lab. Fundam. Study, Tomsk State
Pedagogical University, Tomsk}}
\email{odintsov@ieec.uab.es}
\affiliation{Instituci\`{o} Catalana de Recerca i Estudis Avan\c{c}ats (ICREA)
and Institut de Ciencies de l'Espai (IEEC-CSIC),
Campus UAB, Facultat de Ciencies, Torre C5-Par-2a pl, E-08193 Bellaterra
(Barcelona), Spain}

\begin{abstract}

We consider class of modified $f(R)$ gravities with the effective
cosmological constant epoch at the early and late universe. Such models pass
most of solar system tests as well they satisfy to cosmological bounds.
Despite their very attractive properties, it is shown that one realistic class of
such models may lead to significant Newton law corrections at large
cosmological scales. Nevertheless, these corrections are small at solar
system as well
as  at the future universe. Another realistic  model with acceptable
Newton law regime shows the matter instability.

\end{abstract}

\pacs{11.25.-w, 95.36.+x, 98.80.-k}

\maketitle

\section{Introduction}

Much attention has been paid recently to the study of modified $f(R)$
gravity (for review, see \cite{review}). Such theory which may be related
with string effective action \cite{string}) may successfully
describe the dark energy epoch \cite{CDTT,NO,FR,FR1,cap,lea}.
It is remarkable that even the form of $f(R)$ gravity may be reconstructed
from the known universe expansion history\cite{recrev}. Hence, this
approach suggests the gravitational alternative for dark energy. It may be
considered as proposal for new gravity theory which could be more exact
than usual General Relativity at current/future universe.  If it is so,
such a theory should pass the known solar system tests \cite{DK,NO,FR,FR1}
as well as cosmological bounds. Unfortunately, despite the significant
progress in the construction of more-less acceptable models, the totally
satisfactory theory has not been yet proposed.

Recently, $f(R)$ gravity with the early/late-time effective cosmological
constant epoch
was proposed \cite{cap,recrev,HS,AB}. The very attractive, simple versions
of such theory \cite{HS,AB} seem to show quite satisfactory behaviour from
the cosmological point of view (the models of ref.\cite{cap,recrev} are
quite complicated). As well they seem to satisfy (most) of  solar system
tests. Nevertheless, some  deviations from General Relativity may be
expected.
Specifically, the model \cite{HS} may show
 large Newton law corrections at cosmological scales. Nevertheless, for
limited range of parameters these corrections are small in Solar System.
As well they  become small at the future
universe. On the same time, the model of ref.\cite{AB} may lead to matter
instability in the proposed range of parameters. This indicates that such
theories which show remarkably beatiful behaviour as $\Lambda$CDM
cosmologies should be extended, perhaps, introducing more parameters.

\section{ The Newton law corrections in f(R) gravity with an effective
cosmological constant epoch}

The action of general $f(R)$ gravity (for a review, see \cite{review}) is
given by
\be
\label{XXX7}
S=\frac{1}{\kappa^2}\int d^4 x \sqrt{-g} \left(R + f(R)\right)\ .
\ee
Here $f(R)$ is an arbitrary function.
The general equation of motion  in $f(R)$-gravity with matter is given by
\be
\label{XXX22}
\frac{1}{2}g_{\mu\nu} F(R) - R_{\mu\nu} F'(R) - g_{\mu\nu} \Box F'(R)  + \nabla_\mu \nabla_\nu F'(R)
= - \frac{\kappa^2}{2}T_{(m)\mu\nu}\ .
\ee
Here $F(R)=R+f(R)$ and $T_{(m)\mu\nu}$ is the matter energy-momentum tensor.

By introducing the auxilliary field $A$ one may rewrite the action
(\ref{XXX7}) in the following form:
\be
\label{XXX10}
S=\frac{1}{\kappa^2}\int d^4 x \sqrt{-g} \left\{\left(1+f'(A)\right)\left(R-A\right) + A + f(A)\right\}\ .
\ee
 From the equation of motion with respect to $A$, it follows $A=R$.
By using the scale transformation $g_{\mu\nu}\to \e^\sigma g_{\mu\nu}$ with $\sigma = -\ln\left( 1 + f'(A)\right)$,
we obtain the Einstein frame action \cite{NO}:
\bea
\label{XXX11}
S_E &=& \frac{1}{\kappa^2}\int d^4 x \sqrt{-g} \left\{ R - \frac{3}{2}\left(\frac{F''(A)}{F'(A)}\right)^2
g^{\rho\sigma}\partial_\rho A \partial_\sigma A - \frac{A}{F'(A)}
+ \frac{F(A)}{F'(A)^2}\right\} \nn
&&=\frac{1}{\kappa^2}\int d^4 x \sqrt{-g} \left( R - \frac{3}{2}g^{\rho\sigma}
\partial_\rho \sigma \partial_\sigma \sigma - V(\sigma)\right)\ , \\
V(\sigma) &=& \e^\sigma g\left(\e^{-\sigma}\right) - \e^{2\sigma} f\left(g\left(\e^{-\sigma}\right)\right)
= \frac{A}{F'(A)} - \frac{F(A)}{F'(A)^2}\ .
\eea
Here $g\left(\e^{-\sigma}\right)$ is given by solving $\sigma = -\ln\left( 1 + f'(A)\right)=\ln F'(A)$
as $A=g\left(\e^{-\sigma}\right)$.
After the scale transformation  $g_{\mu\nu}\to \e^\sigma g_{\mu\nu}$,
there appears a coupling of the scalar field $\sigma$
with the matter. For example, if the matter is the scalar field $\Phi$ with mass $M$, whose action is given by
\be
\label{MN1}
S_\phi=\frac{1}{2}\int d^4x\sqrt{-g}\left(-g^{\mu\nu}\partial_\mu\Phi \partial_\nu\Phi - M^2 \Phi^2\right)\ ,
\ee
there appears a coupling with $\sigma$ in the Einstein frame:
\be
\label{MN2}
S_{\phi\, E}=\frac{1}{2}\int d^4x\sqrt{-g} \left(-\e^{\sigma} g^{\mu\nu}\partial_\mu\Phi \partial_\nu\Phi
 - M^2 \e^{2\sigma}\Phi^2\right)\ .
\ee
The strength of the coupling is  of the gravitational coupling $\kappa$
order. Unless the mass of $\sigma$,
which is defined by
\be
\label{MN3}
m_\sigma^2 \equiv \frac{1}{2}\frac{d^2 V(\sigma)}{d\sigma^2}
\ee
is large, there appears the large correction to the Newton law.

More exactly,
in the Einstein frame, matter fields give a source term for the scalar field $\sigma$ like
\be
\label{SS1}
J_\sigma = \e^{a\sigma}\rho\ .
\ee
Here $\rho$ is the energy density (in the Jordan frame). Now we consider the fluctuations from
the background of $\sigma=\sigma_0$ ($\sigma_0$ is not always a constant):
\be
\label{SS2}
\sigma=\sigma_0 + \delta \sigma\ .
\ee
For simplicity, we consider the limit where the spacetime is almost flat and consider the point like souces
\be
\label{SS3}
J_\sigma = \rho_{0}^{(1)}\e^{a^{(1)}\sigma(x)}\delta\left(x-x^{(1)}\right)
+ \rho_{0}^{(2)}\e^{a^{(2)}\sigma(x)}\delta\left(x-x^{(2)}\right)\ .
\ee
Then by the propagation of $\delta\sigma$, we find the following correlation function
\be
\label{SS4}
\left<\e^{a^{(1)}\sigma(x^{(1)})}\e^{a^{(1)}\sigma(x^{(2)})}\right>
\sim \e^{\left(a^{(1)} + a^{(2)}\right)\sigma_0 + a^{(1)}a^{(2)}G_\sigma(x_1,x_2)}\ .
\ee
Here $G_\sigma(x_1,x_2)$ is the correlation function of $\sigma$. When the mass of $\sigma$ is small, we have
\be
\label{SS5}
G_\sigma(x_1,x_2)=\frac{\kappa^2}{12\pi (x^{(1)} - x^{(2)})^2}\ ,
\ee
and
\be
\label{SS6}
\left<\e^{a^{(1)}\sigma(x^{(1)})}\e^{a^{(1)}\sigma(x^{(2)})}\right>
\sim \e^{\left(a^{(1)} + a^{(2)}\right)\sigma_0 + \frac{a^{(1)}a^{(2)}\kappa^2}{12\pi (x^{(1)} - x^{(2)})^2}}\ .
\ee
At the long range where $(x^{(1)} - x^{(2)})^2$ is large enough compared
with $\kappa^2$, we find
\be
\label{SS7}
\left<\e^{a^{(1)}\sigma(x^{(1)})}\e^{a^{(1)}\sigma(x^{(2)})}\right>
\sim \e^{\left(a^{(1)} + a^{(2)}\right)\sigma_0}\left( 1 + \frac{a^{(1)}a^{(2)}\kappa^2}{12\pi (x^{(1)} - x^{(2)})^2}
+ \cdots \right)\ .
\ee
Then there appears the long range force and the strength of the coupling
is given by
$\e^{\left(a^{(1)} + a^{(2)}\right)\sigma_0}a^{(1)}a^{(2)}\kappa^2$\ .
If the coupling is very small, the correction to the Newton law might
 be not so small.

Recently very interesting $f(R)$ model has been proposed by Hu and Sawicki
\cite{HS}. In the model $f(R)$ is given by
\be
\label{HS1}
f_{HS}(R)=-\frac{m^2 c_1 \left(R/m^2\right)^n}{c_2 \left(R/m^2\right)^n + 1}\ ,
\ee
which satisfies the condition
\bea
\label{HS2}
\lim_{R\to\infty} f_{HS} (R) &=& \mbox{const}\ ,\nn
\lim_{R\to 0} f_{HS}(R) &=& 0\ ,
\eea
The estimation of ref.\cite{HS} suggests that $R/m^2$ is not so small but
rather large even in the present
universe and $R/m^2\sim 41$. Then we have
\be
\label{HSb1}
f_{HS}(R)\sim - \frac{m^2 c_1}{c_2} + \frac{m^2 c_1}{c_2^2} \left(\frac{R}{m^2}\right)^{-n}\ ,
\ee
which gives an ``effective'' cosmological constant $-m^2 c_1/c_2$.
 The effective cosmological
constant generates the accelerating expansion in the present universe.
Then
\be
\label{HSbb1}
H^2 \sim \frac{m^2 c_1 \kappa^2 }{c_2} \sim \left(70 \rm{km/s\cdot pc}\right)^2 \sim \left(10^{-33}{\rm eV}\right)^2\ .
\ee
In the intermediate epoch, where the matter density $\rho$ is larger
than the effective cosmological constant,
\be
\label{HSbb2}
\rho > \frac{m^2 c_1}{c_2}\ ,
\ee
there appears the matter dominated phase (such phase may occur for other
modified $f(R)$ gravity as well \cite{cap,lea}) and the universe could
expand with deceleration. Hence, above model leads to the effective
$\Lambda$CDM cosmology like models \cite{cap,recrev}.

Some remark is in order. The approximate
expression for the Hu-Sawicky model should be taken with great care.
The reason is that at very small curvatures where the (non-perturbative)
 function $f(R)$ goes to zero, the approximation breaks down (the
corresponding function $f$ may become singular).

Due to the scalar field in (\ref{XXX11}), an extra (fifth) force
could manifest itself. It
could violate the Newton law. The Newton law is well understood and
its correction should be very small at least in the present universe.
If the mass of $\sigma$ is large enough in the present universe,
the problem could be avoided. We now investigate the model by assuming
$A/m^2=R/m^2 \gg 1$ since $R/m^2 \sim 41$ even in the present universe.
Then one gets
\be
\label{HSb2}
\sigma \sim - \frac{n c_1}{c_2}\left(\frac{A}{m^2}\right)^{-n-1}\ ,\quad
V(\sigma) \sim \frac{m^2 c_1}{c_2} - \frac{(n+1)m^2 c_1}{c_2^2}\left(\frac{A}{m^2}\right)^{-n}\ ,
\ee
and
\be
\label{HSb3}
m_\sigma^2 \equiv \frac{1}{2}\frac{d^2 V(\sigma)}{d\sigma^2}
=- \frac{1}{2}\left(\frac{d\sigma}{dA}\right)^{-3}\frac{d^2\sigma}{dA}\frac{dV}{dA}
+ \frac{1}{2}\left(\frac{d\sigma}{dA}\right)^{-2}\frac{d^2 V}{dA^2}
=\frac{1}{2}\left\{\frac{A}{F'(A)} - \frac{4F(A)}{\left(F'(A)\right)^2} + \frac{1}{F''(A)}\right\}
\sim \frac{m^2}{2nc_1}\left(\frac{A}{m^2}\right)^{n+2}\
\ee
First we consider  the universe at very large scales, where $R\sim
\left(10^{-33}\,{\rm eV}\right)^{-2}$ and
therefore $R/m^2\sim 41$. If $c_1$ is not so small and/or $n$ is not so
large,
$R/m^2\sim 41$, we find $m_\sigma$ should be very small $m_\sigma \sim 10^{-33}\,{\rm eV}$.
Therefore, the correction to the Newton law is large.
Note that $\sigma_0\sim 0$ in (\ref{SS7}) for the model \cite{HS}.
Since $a_{1,2}\sim 1$,  the correction to the Newton law could  be not so small.

Although $m_\sigma$ could be very small at large scales since
$R_0$ is very small,
$R_0$ can be larger near or in the star.
Since $1\,{g}\sim 6\times 10^{32}\,{\rm eV}$ and $1\,{\rm cm}\sim \left(2\times 10^{-5}\,{\rm eV}\right)^{-1}$,
the density is about $\rho\sim 1 {\rm g/cm^3} \sim 5\times 10^{18}\, {\rm
eV}^4$ inside the earth.
This shows that the magnitude of the curvature could be $R_0 \sim \kappa^2
\rho \sim \left(10^{-19}\,{\rm eV}\right)^2$
and therefore $R_0/m^2 \sim 10^{28}$. Hence, in case $n=2$, we find
$m_\sigma\sim 10^{19}\,{\rm GeV}$, which is
very large and the correction to the Newton law is very small.

Even in air, one finds $\rho\sim 10^{-6} {\rm g/cm^3} \sim 10^{12}\, {\rm eV}^4$, which gives
$R_0 \sim \kappa^2 \rho \sim \left(10^{-25}\,{\rm eV}\right)^2$ and $R_0/m^2 \sim 10^{16}$.
In case $n=2$,  $m_\sigma\sim 10^{-1}\, {\rm eV}$, which gives
a correlation length (Compton wave length)
about $1\,\mu{\rm m}$. Thus, the correction to the Newton law could not be
observed on the earth for such a model.

What happens in the solar system?
In the solar system, there could be interstellar gas. Typically, in the
interstellar gas, there is one proton
(or hydrogen atom) per $1\,{\rm cm}^3$, which shows $\rho\sim 10^{-5}\,
{\rm eV}^4$, $R_0\sim 10^{-61}\, {\rm eV}^2$,
and therefore $R_0/m^2 \sim 10^4$. Then for $n=2$, we find $m_\sigma \sim 10^{-25}\,{\rm eV}$,
which corresponds to the correlation length of $10^{18}\,{\rm m}\sim 100\, {\rm pc}$.
Then the correction to the Newton law could be observed.
In case $n=8$, however, we find $m_\sigma \sim 10^{-13}\,{\rm eV}$,
which corresponds to the correlation length of $10^{6}\,{\rm m}$, which is
less than the radius of earth ($\sim 10^7\,{\rm m}$).
Then the correction to the Newton law could not be observed.
Hence, some sub-class of above theory may pass known solar sytem tests at
the scales of the solar system order.


In (\ref{HSb3}),  the Einstein frame was considered (\ref{XXX11}).
However, similar conclusions may be made also in Jordan frame.
By multipling (\ref{XXX22})  with $g^{\mu\nu}$, one obtains
\be
\label{SS12}
 - 3 \Box F'(R) - R F'(R) + 2 F(R) = - \frac{\kappa^2}{2} T\ .
\ee
Here $T\equiv T_{(m)\rho}^{\ \rho}$. The equation (\ref{SS12}) corresponds to Eq.(39) in \cite{HS}.
Now we consider the background where $R$ is a constant $R=R_0$, that is, (anti-)de Sitter space
which can be obtained by solving the algebraic equation
\be
\label{SS13}
- R_0 F'(R_0) + 2 F(R_0) = 0\ .
\ee
Since $\e^{-\sigma} = F'(R)$, one gets
\be
\label{SS14}
\delta R = - \frac{F'(R)}{F''(R)} \delta\sigma \ .
\ee
Consider the fluctuation
\be
\label{SS15}
R=R_0 + \delta R\ ,
\ee
which leads to
\be
\label{SS16}
\Box \delta\sigma - \frac{1}{3}\left(\frac{F'(R_0)}{F''(R_0)} - R_0\right) \delta\sigma = - \frac{\kappa^2}{6F'(R_0)}T\ .
\ee
One may consider the point source
\be
\label{SS17}
T=T_0 \delta (x)\ .
\ee
Then the solution of (\ref{SS16}) is given by
\be
\label{SS18}
\delta\sigma = - \frac{\kappa^2 T_0}{6F'(R_0)}G(m^2,|x|)\ .
\ee
Here
\be
\label{SS19}
m^2\equiv \frac{1}{3}\left(\frac{F'(R_0)}{F''(R_0)} - R_0\right) \ ,\quad
\left(\Box - m^2\right)G(m^2,|x|)= \delta(x)\ .
\ee
If $m^2<0$, there appears tachyon and there could be some instability.
Even if $m^2>0$, when $m^2$ is small, $\delta R\neq 0$ at long ranges,
which generates the large correction to the Newton law.
In case of \cite{HS}, we find, when $R/m^2 \gg 1$ as in the present universe,
\be
\label{SS20}
m^2 \sim \frac{m^2 c_2^2}{3n(n+1)c_1}\left(\frac{R_0}{m^2}\right)^{n+2}\ .
\ee
Compared this expression (\ref{SS20}) with (\ref{HSb3}) by putting $A=R_0$, we find
$m^2\sim m_\sigma^2$. Then the correction to the Newton law
is the same.


In \cite{HS}, it is assumed  $R/m^2\gg 1$ but it might be interesting to
study the model  assuming $R/m^2\ll 1$, which may correspond to the
future universe.
When $A/m^2=R/m^2\ll 1$, the potential $V(\sigma)$  (\ref{XXX11}) is given by
\be
\label{HS3}
V(\sigma) \sim \left(1-n\right)c_1 m^2 \left(A/m^2\right)^n\ ,
\ee
and we find
\be
\label{HS4}
\sigma \sim - \ln \left(1 - nc_1 \left(A/m^2\right)^{n-1}\right)\ .
\ee
Let us consider the case $n>1$ and $0<n<1$ separately.

In case $n>1$, when $A$ is small, (\ref{HS4}) can be written as
\be
\label{HS5}
\sigma \sim n c_1 \left(A/m^2\right)^{n-1}\ ,
\ee
and therefore
\be
\label{HS6}
V(\sigma) \sim \left(1-n\right)c_1 m^2 \left(\frac{\sigma}{n c_1}\right)^{n/(n-1)}\ .
\ee
Then
\be
\label{HS7}
m_\sigma^2 \sim \frac{n-2}{2n(n-1)c_1} \left(\frac{\sigma}{n c_1}\right)^{-1 + 1/(n-1)}\ .
\ee
Note $m_\sigma>0$ if $c_1>0$.
Eq.(\ref{HS5}) shows that $\sigma$ is small when $A/m^2$ is small. Then
the mass $m_\sigma$ becomes large
when the curvature $R\sim A$. Therefore the scalar field does not propagate
 at large
ranges and the Newton law could not be violated.

On the other hand, in case $n<1$, for small $A$, we find
\be
\label{HS8}
\sigma \sim - (n-1)\ln \frac{A}{m^2} + \ln(-nc_1)\ .
\ee
When $A$ is small, $\sigma$ is negative and large. Eq.(\ref{HS8}) shows
\be
\label{HS9}
V(\sigma)\sim \left(1-n\right)c_1 m^2 \left(-nc_1\right)^{n/(n-1)}\e^{-n\sigma/(n-1)}\ .
\ee
In order that the potential being real, $c_1$ should be negative. Since
\be
\label{HS10}
m_\sigma^2 \sim \frac{n^2 c_1 m^2 }{1-n}\left(-nc_1\right)^{n/(n-1)}\e^{-n\sigma/(n-1)}\ ,
\ee
the squared mass $m_\sigma^2$ is negative since $c_1<0$, which shows that
$\sigma$ is tachyon and unstable.
Tachyon is inconsistent with quantum theory. Classically if we consider
the perturbation with respect to $\sigma$,
the perturbation becomes large. Since $\sigma$ is related with the curvature by $\sigma=-\ln F'(A) = - \ln F'(R)$,
the instability may indicate the solution where
 by the perturbation, the
curvature of the universe could become large.

Hence, it seems there may be significant correction to Newton law in the
$f(R)$ gravity model under consideration at cosmological scales. It is
remarkable that such
correction becomes negligible in the future, at least, for some
range of parameters.

\section{The absence of matter instability}

There may exist another type of instability (so-called matter
instability) in $f(R)$ gravity \cite{DK}.
The example of the model without such instability is given in \cite{NO}
(for related discussions of matter instability, see, \cite{Faraoni}).
Let us show that current and related models are free from such
instability.
The instability might occur when the curvature is rather large,
 as on the planet, compared with the average curvature
in the universe $R\sim \left(10^{-33}\,{\rm eV}\right)^2$.
By multipling Eq.(\ref{XXX22}) with
$g^{\mu\nu}$, one obtains
\be
\label{XXX23}
\Box R + \frac{F^{(3)}(R)}{F^{(2)}(R)}\nabla_\rho R \nabla^\rho R
+ \frac{F'(R) R}{3F^{(2)}(R)} - \frac{2F(R)}{3 F^{(2)}(R)}
= \frac{\kappa^2}{6F^{(2)}(R)}T\ .
\ee
Here $T\equiv T_{(m)\rho}^{\ \rho}$.
We consider a perturbation from the solution of the Einstein gravity:
\be
\label{XXX24}
R=R_0\equiv - \frac{\kappa^2}{2}T>0\ .
\ee
Note that $T$ is negative since $|p|\ll \rho$ on the earth and
$T=-\rho + 3 p \sim -\rho$. Then we assume
\be
\label{XXX25}
R=R_0 + R_1\ ,\quad \left(\left|R_1\right|\ll \left|R_0\right|\right)\ .
\ee
Now one can get
\bea
\label{XXX26}
0&=&\Box R_0 + \frac{F^{(3)}(R_0)}{F^{(2)}(R_0)}\nabla_\rho R_0 \nabla^\rho R_0 + \frac{F'(R_0) R_0}{3F^{(2)}(R_0)}
 - \frac{2F(R_0)}{3 F^{(2)}(R_0)} - \frac{R_0}{3F^{(2)}(R_0)} \nn
&& + \Box R_1 + 2\frac{F^{(3)}(R_0)}{F^{(2)}(R_0)}\nabla_\rho R_0 \nabla^\rho R_1 + U(R_0) R_1\ , \nn
U(R_0)&\equiv& \left(\frac{F^{(4)}(R_0)}{F^{(2)}(R_0)} - \frac{F^{(3)}(R_0)^2}{F^{(2)}(R_0)^2}\right)
\nabla_\rho R_0 \nabla^\rho R_0 + \frac{R_0}{3} \nn
&& - \frac{F^{(1)}(R_0) F^{(3)}(R_0) R_0}{3 F^{(2)}(R_0)^2} - \frac{F^{(1)}(R_0)}{3F^{(2)}(R_0)}
+ \frac{2 F(R_0) F^{(3)}(R_0)}{3 F^{(2)}(R_0)^2} - \frac{F^{(3)}(R_0) R_0}{3 F^{(2)}(R_0)^2}
\ .
\eea
If $U(R_0)$ is positive, since $\Box R_1 \sim - \partial_t^2 R_1$,
the perturbation $R_1$ is exponentially large and the system
becomes unstable. One may regard $\nabla_\rho R_0\sim 0$ if it is assumed
the matter is
almost uniform as inside the earth.

For the model (\ref{HS1}), by assuming $R_0/m^2\gg 1$, it follows
\be
\label{HS11}
U(R_0)\sim - \frac{m^2 c_2^2}{3c_1 n (n+1)} \left(R_0/m^2\right)^{n+2}\ ,
\ee
which is large and negative if $c_1>0$. Hence, there is no instability
in the sense of ref.\cite{DK}.
When $c_1<0$, however, there could be an instability.
In first ref. of \cite{Faraoni}, a simple condition for the stability in
a sense of
\cite{DK} was given, that is, theory is stable if $F''(R_0)=f''(R_0)>0$
but
unstable if $F''(R_0)=f''(R_0)<0$. Now we have
\be
\label{HS11B}
F''(R_0)\sim \frac{n(n+1)m^2 c_1}{c_2^2}\left(\frac{R_0}{m^2}\right)^{-n-2}\ .
\ee
Then $F''(R_0)\sim -1/U(R_0)>0$ if $c_1$ is positive and
theory is not stable.

As one more example satisfing the conditions (\ref{HS2}), we now consider
\be
\label{HS12}
f_A(R)= -\frac{m^2 c_1}{c_2}\left( 1 - \e^{- \frac{c_2\left(R/m^2\right)^n}{c_2 \left(R/m^2\right)^n + 1}}
\right)\ ,
\ee
The asymptotic behaviors of (\ref{HS12}) are identical with the model (\ref{HS1}) when $R$ is large,
\be
\label{HS14}
f_A(R) \sim f_{HS}(R) \to -\frac{m^2 c_1}{c_2}\ ,
\ee
and when $R$ is small
\be
\label{HS15}
f_A(R) \sim f_{HS}(R) \to - m^2 c_1 \left(R/m^2\right)^n\ .
\ee
Then asymptotic behaviors of the universe does not change and the correction to the Newton law could be large when $R$ is large
and small when $R$ is small. The instability is also absent, as one
can reobtain the results identical with (\ref{HS3}-\ref{HS11}).

Another example is
\be
\label{HS13}
f_B(R)= - f_0 \e^{-\frac{\tilde{m^4}}{R^2}}\ ,
\ee
with a positive constants $f_0$ and $m^4$.
As in the model (\ref{HS1}) in \cite{HS}, we may assume $R/m^2 \gg 1$ from the early universe to the present universe.
Even in the model (\ref{HS14}), we assume $R^2\gg {\tilde m}^4$.
Then by expanding $f_B(R)$ with respect to $m^4/R^2$, we find
\be
\label{HS13b}
f_B(R)\sim - f_0 + f_0 \frac{{\tilde m}^4}{R^2}\ .
\ee
By comparing (\ref{HS13b}) with (\ref{HSb1}), we may identify
\be
\label{HS13c}
n\leftrightarrow 2\ ,\quad f_0\leftrightarrow \frac{m^2 c_1}{c_2}\ ,\quad
f_0{\tilde m}^4 \leftrightarrow \frac{m^6 c_1}{c_2^2}\ .
\ee
Hence, $f_0$ plays the role of the cosmological constant if $f_0>0$
\be
\label{HS13c2B}
H^2 \sim f_0 \sim \left(70 \rm{km/s\cdot pc}\right)^2 \sim \left(10^{-33}{\rm eV}\right)^2\ .
\ee
Thus, the accelerated expansion of the present universe could be generated
by the effective cosmological constant $f_0$.
As in (\ref{HSbb2}), in the earlier but not primordial universe, the
matter density $\rho$ is larger than
the effective cosmological constant $f_0$.
Hence, there occurs the matter dominated phase and the universe could
have
expanded with deceleration.
The aymptotic behavior when the curvature is large is identical with the model (\ref{HS1}), the correction to the
Newton law could  be not so small.

We now investigate also the case that the curvature is small.
Then for the model (\ref{HS14}), we obtain
\be
\label{HS16}
V(\sigma)=\left( - \frac{A^4}{2f_0 {\tilde m}^4} + \frac{A^6}{4f_0 {\tilde m}^8}\right)\e^{-\frac{{\tilde m}^4}{A^2}}\ ,\quad
\sigma \sim \frac{2f_0 {\tilde m}^4}{A^3}\e^{-\frac{{\tilde m}^4}{A^2}}\ .
\ee
and
\be
\label{HS17}
\frac{d^2V(\sigma)}{d\sigma^2}=- \left(\frac{d\sigma}{dA}\right)^{-3}\frac{d^2\sigma}{dA^2}\frac{dV}{dA}
+ \left(\frac{d\sigma}{dA}\right)^{-2}\frac{d^2 V}{dA^2}
\sim \frac{A^{10}}{416 {\tilde m}^{12} f_0^3}\e^{m^4/A^2}\ .
\ee
If $f_0$ is positive, $m_\sigma\equiv (1/2)(d^2V/d\sigma^2)$ is positive
and large when the curvature $R=A$
is small and therefore there is no large corrrection to the Newton law.
We should note, however, if $f_0$ is negative, which corresponds to the model in (\ref{HS1}) $m_\sigma^2$ becomes negative
and there could occur an instability.
On the other hand, when the curvature is large, $U(R_0)$ in (\ref{XXX26}) has the following form:
\be
\label{HS18}
U(R_0)\sim - \frac{R_0^4}{18 f_0 {\tilde m}^4}R_0\ ,
\ee
which is negative and large and therefore there is no instability.
In fact, since
\be
\label{HS18B}
f_B''(R_0)\sim \frac{6f_0 {\tilde m}^4}{R_0^4}>0\ ,
\ee
the condition from first ref. of \cite{Faraoni} is satisfied.

Recently another interesting $f(R)$ model was proposed in \cite{AB}, where
\be
\label{HS19}
F_{AB}(R)=R+f_{AB}(R)
=\frac{R}{2} + \frac{1}{2a}\ln \left[\cosh (aR) - \tanh(b) \sinh(aR)\right]\ ,
\ee
with positive constants $a$ and $b$ (for first $f(R)$ models with
log-terms, see first ref. in \cite{FR1}).

Since the correction to the Newton law has been studied in \cite{AB},
we now investigate the possible instability  for the model (\ref{HS19}).
Since
\be
\label{HS19B}
F''(R)=2a \frac{\left(1-\tanh(b)\right)}{\left(1 + \tanh(b)\right)} \e^{-2aR}\ ,
\ee
it is positive. Then the condition  \cite{Faraoni} seems to be satisfied
and therefore theory seems to be consistent.

When the curvature $R$ is large, one finds
\be
\label{HS20}
F_{AB}(R)\sim R + \frac{1}{2a}\ln\frac{1 - \tanh(b)}{2}
+ \frac{\left(1+\tanh(b)\right)\e^{-2aR}}{2a\left(1-\tanh(b)\right)} + {\cal O}\left(\e^{-4aR}\right)\ .
\ee
Then $U(R_0)$  (\ref{XXX26}) has the following form:
\be
\label{HS21}
U(R_0)\sim - \frac{\e^{2aR_0}}{6a}\frac{\left(1-\tanh(b)\right)}{\left(1 + \tanh(b)\right)}
\left( 1 + 2 \ln\frac{1 - \tanh (b)}{2}\right)\ .
\ee
If $1 + 2 \ln \left(\left(1 - \tanh (b)\right)/2\right)>0$, $U(R_0)$ is very large and negative
and therefore there is no instability.
In \cite{AB}, $b$ is choosen to be $b\gtrsim 1.2$, so
\be
\label{HS22}
1 + 2 \ln\frac{1 - \tanh (1.2)}{2}=-3.97<0\ ,
\ee
and therefore the matter instability  seems to occur.
This indicates that such model should be considered in the other range of
parameters.

\section{Discussion}

In the present letter we considered some solar system tests for several
modified gravities which satisfy to conditions (16).
These theories show very realistic cosmological behaviour and may
easily lead to $\Lambda$CDM cosmology.  It is shown that the theory (15)
  passes known solar system tests as well as
cosmological bounds. Signficant Newton law corrections appear only beyond
the solar system scales as well as for specific values of curvature power
which puts some bound for such theory.
Theory (59) which has an acceptable Newton law regime
shows the matter instability in
the proposed range of the parameters. Thus, the suggested class of models
seems to be very realistic and looks like the alternative for
$\Lambda$CDM. More
accurate and detailed check of cosmological bounds for
such theories should be done but in any case it is expected that some
(combination/extension) of such theories may fit with observable
cosmological data.

\section*{Acknowledgements}

We thank M. Sasaki and W. Hu for useful discussions.
The investigation by S.N. has been supported in part by the
Ministry of Education, Science, Sports and Culture of Japan under
grant no.18549001
and 21st Century COE Program of Nagoya University
provided by Japan Society for the Promotion of Science (15COEG01),
and that by S.D.O. has been supported in part  by the
projects FIS2006-02842, FIS2005-01181
(MEC,Spain), by the project 2005SGR00790 (AGAUR,Catalunya), by LRSS
project N4489.2006.02  and by RFBR grant 06-01-00609 (Russia).

\end{document}